\documentclass[%
 aip,
 jmp,%
 amsmath,amssymb,
 reprint,%
]{revtex4-1}

\usepackage{graphicx}
\usepackage{dcolumn}
\usepackage{bm}

\usepackage[cal=boondox]{mathalfa}

\DeclareMathAlphabet\mathbfcal{OMS}{cmsy}{b}{n}

\usepackage{fixdif}
\DeclareMathOperator{\Sp}{Sp}
\DeclareMathOperator{\Frac}{Frac}
\DeclareMathOperator{\sgn}{sgn}

\begin{document}

\preprint{AIP/123-QED}

\title[Linear Coupling of Transverse Betatron Motion...]{Linear Coupling of Transverse Betatron Oscillations. Dynamic Stability and Invariants of Motion}

\author{Stephan I. Tzenov}
\email{stephan@zjlab.ac.cn}
\affiliation{Zhangjiang Laboratory, 99 Haike Rd, Pudong New District, Shanghai, China}

\author{Zhichu Chen}%
\affiliation{Shanghai Advanced Research Institute (SARI), 99 Haike Rd, Pudong New District, Shanghai, China}%

\author{Hailong Wu}%
\affiliation{Shanghai Advanced Research Institute (SARI), 99 Haike Rd, Pudong New District, Shanghai, China}%


\date{\today}

\begin{abstract} 

Based on the technique of the discrete one-turn transfer maps, the problem of linear coupling between horizontal and vertical betatron oscillations in an accelerator has been treated exactly and entirely in explicit form. The stability region in the fractional part of the horizontal and the vertical betatron tune space as a function of the linear coupling strength, has been obtained, and the increment/decrement of the horizontal and the vertical betatron oscillations in the case of the linear sum resonance has been shown to be approximately equal to the half of the coupling strength. 

The normal form parameterization of the one-turn linear map with horizontal-to-vertical coupling has been developed in detail in the spirit of the Edwards and Teng formalism. The motion in the normal mode in the new normal form coordinates is decoupled implying that two independent Courant-Snyder invariants exist, which have been found explicitly.

\end{abstract}

\pacs{29.20.Dh, 29.20.db, 29.27.Bd}
\keywords{Storage rings and colliders, Linear coupling, Beam dynamics, Canonical transformation}
\maketitle

%

\section{\label{sec:intro}Introduction} 

Linear optics in accelerator rings and transfer lines in the case of uncoupled transverse directions are conventionally described by means of the Twiss parameters. The Twiss parameters or lattice functions are on one hand purely determined by the magnetic structure of the machine or the transfer line (a sort of a device hardware). On the other hand, they relate the beam distribution
in phase space at any point along the beam trajectory in an accelerator device to conserved quantities that are properties of the traveling bunch.

In some accelerator applications the analysis of coupled betatron motion is an important part of the machine design. Initially betatron coupling in the transverse plane was perceived as an undesired effect and corresponding efforts were dedicated to suppress it. However, over the recent two decades it was realized that betatron coupling possesses some interesting and useful features and has become an indispensable part of many accelerator proposals. It was gradually realized that the coupling between the two transverse directions can be of considerable practical importance \cite{Barnard,Cai}. One of the most interesting and promising proposals in that direction is the so-called M\"{o}bius scheme \cite{Talman}. A lattice insert is constructed such that it exchanges the horizontal and vertical betatron oscillations according to the rule ${\left( x \longrightarrow z, \; p_x \longrightarrow p_z, \; z \longrightarrow -x, \; p_z \longrightarrow -p_x \right)}$. The effect of exchange between the transverse degrees of freedom can be achieved by placing a solenoid with an integrated solenoid rotation angle equal to $\pi / 2$ [see Eqs. (\ref{RotCanVar1}) and (\ref{RotCanVar2}) below]. When such an insert is added to an ordinary uncoupled accelerator lattice, horizontal betatron motion on one turn becomes vertical on the next turn and vice versa.

In the current literature on accelerator physics, the most frequently used and the best known are two different basic representations. The first parameterization was proposed by Edwards and Teng \cite{Teng,Edwards}, while the second one by Mais and Ripken \cite{Mais,Ripken}. 

The first approach introduced by Edwards and Teng and further developed and worked out in more detail by others \cite{Sagan,Luo,Calaga} consists in defining a sort of a decoupling transformation that puts the $4 \times 4$ transfer matrix into block-diagonal form. Although this technique has some disadvantages like the fact that the lattice functions are not directly related to the beam sizes, and the procedure cannot be easily generalized to more than two degrees of freedom, in our opinion, it is the most elegant and intuitively direct way to describe the coupled betatron motion in particle accelerators and storage rings.

The basic idea of the second approach is to find a transformation from the eigenvectors of the transfer matrix, that puts the transfer matrix into normal form. In other words, the transfer matrix is transformed into a pure rotation. The lattice functions are defined in terms of elements of the normal transform \cite{Mais,Ripken,Lebedev}. The number of lattice functions used to describe the beam optics is usually minimized, and as a result the interpretation of some of these functions is not as simple as one would like them to be. 

There exists yet another representation which is less known than the above two, and that is the parameterization proposed by Qin and Davidson \cite{Qin,Davidson}. Their generalized Courant-Snyder theory \cite{Courant} provides a new parameterization for the 4D symplectic transfer matrix. In particular, all of the quantities of physical importance in the original one-degree-of-freedom Courant-Snyder theory , including the envelope function, envelope equation, phase advance, transfer matrix, and the Courant-Snyder invariant, are generalized to the case of coupled transverse two-degree-of-freedom dynamics. Thus, the envelope function is generalized to a $2 \times 2$ envelope matrix, and the envelope equation is generalized to a matrix envelope equation. 

The Courant-Snyder theory \cite{Courant} for two-dimensional coupled linear optics can be formulated on the basis of the real representation of the Dirac matrices \cite{Baumgarten}. Any real $4 \times 4$ matrix can be expressed as a linear combination of the real Dirac matrices, which allows symplectic transformations in two dimensions to be conveniently represented in therms the fifteen real Dirac matrices (plus the unit symplectic matrix).

In the present article, we develop a description of coupled linear transverse betatron motion that addresses issues inherent in previous approaches, being in spirit closer to the Edwards and Teng formalism. Here, the problem of linear coupling between horizontal and vertical betatron oscillations in an accelerator is treated exactly and entirely in explicit form by means of transfer maps. The subsequent two Sections \ref{sec:basic} and \ref{sec:map} are devoted to the establishment of the main starting points of our further analysis, as well as to the formal inference of the linear transfer map. Since the dynamical effect of one of the sources of linear coupling, the longitudinal solenoid field, can be transformed away as a regular rotation with suitably chosen angle, we can consider that the most general form of the linear coupling between transverse degrees of freedom is set only by the quadrupoles (normal and skew ones with effective strengths). The stability properties of the linear map are analysed in Section \ref{sec:stability}. A new stability diagram of betatron motion with linear coupling between the transverse degrees of freedom in the fractional part of the tune ${\left( \nu_x, \nu_z \right)}$-space as a function of the coupling strength has been presented. The normal form parameterization of the one-turn map is worked out in detail in Section \ref{sec:norform} and Appendices \ref{app:sympmat} and \ref{app:matrix12}. Since the motion in the normal mode is decoupled, there exist two independent Courant-Snyder invariants, which have been found explicitly. The supporting numerical proofs of the analytical results concerning the normal form representation and the existence of the two independent invariants are presented in Section \ref{sec:simulation}. Finally, in Section \ref{sec:conclrem} our conclusions and outlook are sketched out. 

\section{\label{sec:basic}Theoretical Model and Basic Equations}

Optimal performance of storage rings in contemporary synchrotron light sources and circular colliders substantially depends on the control and easy manipulation of the coupling between the transverse degrees of freedom. Characterizing the coupling in a straightforward fashion becomes particularly important when the machine lattice includes regions where betatron motion is coupled by design, as in the solenoid field of the interaction region of a collider, for instance. Let us begin by writing the Hamiltonian governing the transverse betatron oscillations in the case, where solenoidal fields and skew quadrupoles are present 
\begin{eqnarray}
{\widehat{H}} = {\frac {R} {2}} {\left[ {\left( {\widehat{p}}_x + {\frac {S {\widehat{z}}} {2}} \right)}^2 + {\left( {\widehat{p}}_z - {\frac {S {\widehat{x}}} {2}} \right)}^2 \right]} \nonumber 
\\ 
+ {\frac {1} {2 R}} {\left( G_x {\widehat{x}}^2 + G_z {\widehat{z}}^2 \right)} + {\frac {g_0} {R}} {\widehat{x}} {\widehat{z}}, \label{Hamiltonian}
\end{eqnarray}
where 
\begin{equation}
S {\left( \theta \right)} = {\frac {q B_{0s} {\left( \theta \right)}} {p_{0s}}}, \label{Solenoid}
\end{equation}
and $B_{0s} {\left( \theta \right)}$ is the longitudinal component of the magnetic field on the axis of the solenoid. The scaled canonical momenta ${\widehat{p}}_{x,z}$ and their canonically conjugate coordinates ${\widehat{x}}$ and ${\widehat{z}}$ are related to the actual coordinates $x$, $z$ and the actual momenta $p_x$, $p_z$ according to the expressions  
\begin{equation}
{\widehat{x}} = x - \eta D, \qquad \qquad {\widehat{z}} = z, \nonumber  
\nonumber
\end{equation}
\begin{equation}
{\widehat{p}}_x = {\frac {p_x} {p_{0s}}} - {\frac {\eta} {R}} {\frac {\d D} {\d \theta}}, \qquad \qquad {\widehat{p}}_z = {\frac {p_z} {p_{0s}}}, \nonumber
\end{equation}
where $D {\left( \theta \right)}$ is the dispersion function of the machine. In addition, the focusing strengths $G_{x,z}$ are given by the expressions 
\begin{equation}
G_x = g_Q + R^2 K^2, \qquad \qquad G_z = - g_Q,  
\label{FocusStreng}
\end{equation}
where $K = q {\left( B_z^{(D)} \right)}_{x,z=0} / p_{0s}$ is the local machine curvature in the dipole magnets, and 
\begin{equation}
g_Q = {\frac {q R^2} {p_{0s}}} {\left( {\frac {\partial B_z^{(Q)}} {\partial x}} \right)}_{x,z=0}, \qquad g_0 = {\frac {q R^2} {p_{0s}}} {\left( {\frac {\partial B_z^{(S)}} {\partial x}} \right)}_{x,z=0}, \label{FocusStreng1}
\end{equation}
is the magnetic field gradient of the quadrupole and the skew quadrupole magnets, respectively. Finally, the azimuthal angle $\theta = s / R$ along the machine circumference is used as an independent variable instead of the path length $s$, where $R$ is the mean machine radius. 

We wish to cancel the coupling between the transverse coordinates and the corresponding canonical momenta introduced by the terms in the square bracket of Eq. (\ref{Hamiltonian}). For that purpose, we apply an orthogonal canonical transformation at an angle $\sigma {\left( \theta \right)}$ explicitly
depending on the ”time” $\theta$, defined by the generating function 
\begin{eqnarray}
F_2 {\left( {\widehat{x}}, {\widehat{z}}, {\widehat{P}}_x, {\widehat{P}}_z; \theta \right)} = {\widehat{P}}_x {\left( {\widehat{x}} \cos \sigma - {\widehat{z}} \sin \sigma \right)} \nonumber 
\\ 
+ {\widehat{P}}_z {\left( {\widehat{x}} \sin \sigma + {\widehat{z}} \cos \sigma \right)}. \label{GenFunction}
\end{eqnarray}
The relation between the old and the new canonical coordinates can be expressed as 
\begin{equation}
{\widehat{x}} = {\widehat{X}} \cos \sigma + {\widehat{Z}} \sin \sigma, \qquad {\widehat{z}} = - {\widehat{X}} \sin \sigma + {\widehat{Z}} \cos \sigma, \label{RotCanVar1}
\end{equation}
\begin{equation}
{\widehat{p}}_x = {\widehat{P}}_x \cos \sigma + {\widehat{P}}_z \sin \sigma, \qquad {\widehat{p}}_z = - {\widehat{P}}_x \sin \sigma + {\widehat{P}}_z \cos \sigma, \label{RotCanVar2}
\end{equation}
It can be easily verified that the new Hamiltonian acquires the form
\begin{equation}
{\widehat{\mathcal{H}}} = {\frac {R} {2}} {\left(  {\widehat{P}}_x^2 + {\widehat{P}}_z^2 \right)} + {\frac {1} {2 R}} {\left( {\widetilde{G}}_x {\widehat{X}}^2 + {\widetilde{G}}_z {\widehat{Z}}^2 \right)} + {\frac {{\widetilde{g}}_0} {R}} {\widehat{X}} {\widehat{Z}}, \label{NewHamil}
\end{equation}
provided the solenoid rotation angle $\sigma$ satisfies the relation 
\begin{equation}
{\frac {\d \sigma} {\d \theta}} = {\frac {R} {2}} S {\left( \theta \right)}. \label{SolenAngle}
\end{equation}
The new focusing and coupling strengths are 
\begin{equation}
{\widetilde{G}}_x = G_x \cos^2 \sigma + G_z \sin^2 \sigma - g_0 \sin 2 \sigma + {\frac {R^2 S^2} {4}}, \label{FocusCouple1}
\end{equation}
\begin{equation}
{\widetilde{G}}_z = G_x \sin^2 \sigma + G_z \cos^2 \sigma + g_0 \sin 2 \sigma + {\frac {R^2 S^2} {4}}, \label{FocusCouple2}
\end{equation}
\begin{equation}
{\widetilde{g}}_0 = {\frac {1} {2}} {\left( G_x - G_z \right)} \sin 2 \sigma + g_0 \cos 2 \sigma. \label{FocusCouple3}
\end{equation}
Note that the last Eq. (\ref{FocusCouple3}) provides an efficient tool to correct linear coupling induced by skew quadrupoles by using solenoid fields and vice versa. It suffices to choose the strength of the solenoid, such that the rotation angle satisfies the relation 
\begin{equation}
\tan 2 \sigma = {\frac {2 g_0} {G_z - G_x}}. \label{SolenCorr}
\end{equation}
Without loss of generality, we shall assume in what follows that the Hamiltonian describing the linear coupling between the transverse degrees of freedom in an accelerator is of the form (\ref{NewHamil}). For the sake of simplicity the tilde signs of the focusing strength and the coupling coefficients will be omitted. 

\section{\label{sec:map}The Linear Map of Coupled Betatron Oscillations}

One may argue that once the Hamiltonian (\ref{NewHamil}) governing the dynamics of a single particle is properly defined, we can formally write the corresponding Hamilton's equations of motion. The latter can be solved in principle with specified initial conditions, which gives us the complete information about the beam. In the majority of cases of practical interest an analytical solution to the equations of motion is a hopeless exercise, so as the necessity of employing numerical methods arises. Since all numerical methods for solving differential equations involve discretization schemes anyway, it is natural to pose the question about the possibility of substitution of the Hamilton's equations of motion with mapping. For that purpose we perform a second canonical transformation specified by the generating function of the second type \cite{TzenovBOOK} 
\begin{equation}
F_2 {\left( {\widehat{X}}, P_x, {\widehat{Z}}, P_z; \theta \right)} = {\frac {{\widehat{X}} P_x} {\sqrt{\beta_x}}} - {\frac {\alpha_x {\widehat{X}}} {2 \beta_x}} + {\frac {{\widehat{Z}} P_z} {\sqrt{\beta_z}}} - {\frac {\alpha_z {\widehat{Z}}} {2 \beta_z}}
, \label{SecondCanTrans}
\end{equation}
relating the old and the new canonical coordinates according to the relations 
\begin{equation}
{\widehat{U}} = U {\sqrt{\beta_u}}, \qquad  {\widehat{P}}_u = {\frac {1} {\sqrt{\beta_u}}} {\left( P_u - \alpha_u U \right)}, \qquad u = {\left( x, z \right)}, \label{OldNewVar}
\end{equation}
where $\alpha_{x,z}$ and $\beta_{x,z}$ are the well-known Twiss parameters. Then, the Hamiltonian (\ref{NewHamil}) acquires the canonical form 
\begin{equation}
{\mathcal{H}} = {\frac {{\dot{\chi}}_x} {2}} {\left( P_x^2 + X^2 \right)} + {\frac {{\dot{\chi}}_z} {2}} {\left( P_z^2 + Z^2 \right)} + {\frac {{\widetilde{g}}_0} {R}} {\sqrt{\beta_x \beta_z}} X Z, \label{CanonHamil}
\end{equation}
where 
\begin{equation}
{\dot{\chi}}_{x,z} = {\frac {\d \chi_{x,z}} {\d \theta}} = {\frac {R} {\beta_{x,z}}}, \label{DiffPhaAdv}
\end{equation}
is the derivative of the corresponding unperturbed phase advances. 

The problem of linear coupling between horizontal and vertical betatron oscillations in an accelerator can be treated exactly by means of an elegant technique involving transfer maps. The equations for the linear coupling map can be written in the form
\begin{equation}
X_{n+1} = X_n \cos \omega_1 + {\left( P_{x,n} - {\mathfrak{C}} Z_n \right)} \sin \omega_1, \label{CoupledMapX}
\end{equation}
\begin{equation}
P_{x,n+1} = - X_n \sin \omega_1 + {\left( P_{x,n} - {\mathfrak{C}} Z_n \right)} \cos \omega_1, \label{CoupledMapPX}
\end{equation}
\begin{equation}
Z_{n+1} = Z_n \cos \omega_2 + {\left( P_{z,n} - {\mathfrak{C}} X_n \right)} \sin \omega_2, \label{CoupledMapZ}
\end{equation}
\begin{equation}
P_{z,n+1} = - Z_n \sin \omega_2 + {\left( P_{z,n} - {\mathfrak{C}} X_n \right)} \cos \omega_2, \label{CoupledMapPZ}
\end{equation}
where 
\begin{equation}
\omega_{1,2} = 2 \pi \nu_{1,2}, \qquad \qquad {\mathfrak{C}} = {\frac {l {\widetilde{g}}_0 {\left( \theta_0 \right)}} {R^2}} {\sqrt{\beta_x {\left( \theta_0 \right)} \beta_z {\left( \theta_0 \right)}}}, \label{FreqandConst}
\end{equation}
and the coupling source with strength ${\widetilde{g}}_0$ is concentrated in a single point $\theta_0$ along the machine circumference. Moreover, $\nu_{1,2}$ are the betatron tunes associated with the uncoupled part of the Hamiltonian (\ref{CanonHamil}). 

\section{\label{sec:stability}Stability Properties of the Linear Map}

From Eqs. (\ref{CoupledMapX}) and (\ref{CoupledMapPX}) we readily obtain
\begin{equation}
X_{n+1} \cos \omega_1 - P_{x,n+1} \sin \omega_1 = X_n, \label{EquationXPX}
\end{equation}
and a similar expression for $Z$ and $P_z$ from Eqs. (\ref{CoupledMapZ}) and (\ref{CoupledMapPZ}), which plugged back into Eqs. (\ref{CoupledMapX}) and (\ref{CoupledMapZ}), respectively, yield
\begin{equation}
X_{n+1} - 2 X_n \cos \omega_1 + X_{n-1} = - {\mathfrak{C}} Z_n \sin \omega_1, \label{SecondOrdX}
\end{equation}
\begin{equation}
Z_{n+1} - 2 Z_n \cos \omega_2 + Z_{n-1} = - {\mathfrak{C}} X_n \sin \omega_2. \label{SecondOrdZ}
\end{equation}
The last two second-order difference equations are easy to solve by the ansatz 
\begin{equation}
X_n = A {\rm e}^{i \Omega n}, \qquad Z_n = B {\rm e}^{i \Omega n}, \qquad \lambda = {\rm e}^{i \Omega}, \label{Ansatz}
\end{equation}
which substituted in Eqs. (\ref{SecondOrdX}) and (\ref{SecondOrdZ}) result in the linear system of equations for the unknown amplitudes $A$ and $B$ 
\begin{equation}
{\left( \lambda^2 - 2 \lambda \cos \omega_1 + 1 \right)} A + B {\mathfrak{C}} \lambda \sin \omega_1 = 0, \label{AmplitudeA}
\end{equation}
\begin{equation}
A {\mathfrak{C}} \lambda \sin \omega_2 + {\left( \lambda^2 - 2 \lambda \cos \omega_2 + 1 \right)} B = 0. \label{AmplitudeB}
\end{equation}
It has a nontrivial solution if its determinant is equal to zero, namely 
\begin{eqnarray}
\lambda^4 - 2 {\left( \cos \omega_1 + \cos \omega_2 \right)} \lambda^3 + {\left( 2 + 4 \cos \omega_1 \cos \omega_2 - \right.} \nonumber 
\\ 
{\left. {\mathfrak{C}}^2 \sin \omega_1 \sin \omega_2 \right)} \lambda^2 - 2 {\left( \cos \omega_1 + \cos \omega_2 \right)} \lambda + 1 = 0. \label{DispersEquat}
\end{eqnarray}
It is clear that if $\lambda_1$ is a certain root of the dispersion equation (\ref{DispersEquat}), then $1 / \lambda_1$ is also a root, which in general is a basic property of the characteristic polynomial of a symplectic matrix \cite{TzenovBOOK}. This observation allows us to write the above Eq. (\ref{DispersEquat}) in alternative form 
\begin{equation}
{\left( \lambda^2 - \mu_1 \lambda + 1 \right)} {\left( \lambda^2 - \mu_2 \lambda + 1 \right)} = 0, \qquad \mu_k = \lambda_k + {\frac {1} {\lambda_k}}, \label{DispersEquat1}
\end{equation}
where $k = 1, 2$. Comparison of the left-hand-sides of Eq. (\ref{DispersEquat}) and (\ref{DispersEquat1}) yields 
\begin{eqnarray}
\mu_1 + \mu_2 = 2 {\left( \cos \omega_1 + \cos \omega_2 \right)}, \nonumber 
\\ 
\mu_1 \mu_2 = 4 \cos \omega_1 \cos \omega_2 - {\mathfrak{C}}^2 \sin \omega_1 \sin \omega_2. \nonumber
\end{eqnarray}
This implies that $\mu_k$ are the roots of the quadratic equation
\begin{eqnarray}
\mu^2 - 2 {\left( \cos \omega_1 + \cos \omega_2 \right)} \mu \nonumber 
\\ 
+ 4 \cos \omega_1 \cos \omega_2 - {\mathfrak{C}}^2 \sin \omega_1 \sin \omega_2 = 0, \label{QuadEqMuk}
\end{eqnarray}
so that 
\begin{eqnarray}
\mu_{1,2} = \cos \omega_1 + \cos \omega_2 \nonumber 
\\ 
\pm {\sqrt{{\left( \cos \omega_1 - \cos \omega_2 \right)}^2 + {\mathfrak{C}}^2 \sin \omega_1 \sin \omega_2}}. \label{Mu12Roots}
\end{eqnarray}
The solutions of the dispersion equation (\ref{DispersEquat1}) can be represented in alternative form according to 
\begin{equation}
\lambda_k = {\frac {\mu_k} {2}} + {\sqrt{{\frac {\mu_k^2} {4}} - 1}}, \qquad \qquad k = 1, 2, \label{Sigma12Roots}
\end{equation}
so that, for the eigenfrequencies $\Omega_1$ and $\Omega_2$ we finally obtain 
\begin{equation}
\Omega_k = \arccos {\left( {\frac {\mu_k} {2}} \right)}, \qquad \qquad k = 1, 2. \label{Lambda12Eigen}
\end{equation}
The motion is stable if $\mu_{1,2}$ given by Eq. (\ref{Mu12Roots}) simultaneously satisfy the conditions 
\begin{equation}
- 2 \leq \mu_{1,2} \leq 2. \label{Stability}
\end{equation}
\begin{figure}
\begin{center} 
\includegraphics[width=8.0cm]{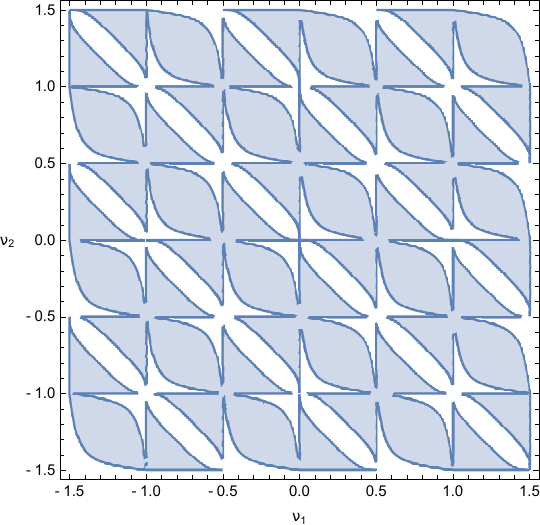}
\caption{\label{fig:stabregion} Stability diagram (the shaded region) of betatron motion with linear coupling between the transverse degrees of freedom in the fractional part of the tune
${\left( \nu_x, \nu_z \right)}$-space. For demonstrativeness, the coupling strength is taken to be ${\mathfrak{C}} = 0.75$.}
\end{center}
\end{figure}
The stability region of betatron oscillations with linear coupling between the transverse degrees of freedom in the fractional part of the betatron tune ${\left( \nu_x, \nu_z \right)}$-space is shown in Fig. \ref{fig:stabregion}. Clearly visible are the instability regions in the vicinity of the linear sum resonances of the form $\Frac{\left( \nu_x \right)} + \Frac{\left( \nu_z \right)} = 0$ and $\Frac{\left( \nu_x \right)} + \Frac{\left( \nu_z \right)} = \pm 1$, where $\Frac{\left( \nu_{x,z} \right)}$ are the fractional parts of the betatron tunes. Details concerning the particular form of the boundaries of the stability diagram are presented in Appendix \ref{app:boundaries}. 

In view of the fact that the sum resonances are significantly more dangerous, let us examine them in more detail. Suppose that the tunes $\nu_1$ and $\nu_2$ satisfy the relation $\nu_1 + \nu_2 = n + \epsilon_s$, where $n$ is an integer and $\epsilon_s$ is the resonance detuning. At exact resonance $\epsilon_s = 0$, from Eqs. (\ref{Mu12Roots}) and (\ref{Lambda12Eigen}) we obtain
\begin{equation}
\mu_{1,2} = 2 \cos \omega_1 \pm i {\mathfrak{C}} \sin \omega_1, \label{ResonMu12}
\end{equation}
\begin{equation}
\Omega_{1,2} = \arccos {\left( \cos \omega_1 \pm {\frac {i {\mathfrak{C}}} {2}} \sin \omega_1 \right)}. \label{ResonLamb12}
\end{equation}
For sufficiently small coupling coefficients ${\mathfrak{C}}$, one can determine approximately the amplitude increment of betatron oscillations at exact linear sum resonance. For the eigentunes we obtain a simple expression 
\begin{equation}
\Omega_{1,2} \approx \omega_1 \mp {\frac {i {\mathfrak{C}}} {2}}. \label{EigenLamb12}
\end{equation}
The above equation shows that the increment of the horizontal and the vertical betatron oscillations is approximately equal to the half of the coupling strength. 

\section{\label{sec:norform}Normal Form Parameterization} 

To compactify notations in what follows, let us introduce the state vector 
\begin{equation}
{\mathbfcal{Z}}_n = 
\begin{pmatrix}
X_n \\
P_{x,n} \\
Z_n \\ 
P_{z,n}
\end{pmatrix}, \label{StateVec}
\end{equation}
and write the linear map (\ref{CoupledMapX}) -- (\ref{CoupledMapPZ}) as 
\begin{equation}
{\mathbfcal{Z}}_{n+1} = {\widehat{\mathbf{G}}} {\mathbfcal{Z}}_n, \label{LinMapMatr}
\end{equation}
where 
\begin{equation}
{\widehat{\mathbf{G}}} = 
\begin{pmatrix}
\cos \omega_1 & \sin \omega_1 & - {\mathfrak{C}} \sin \omega_1 & 0 \\
- \sin \omega_1 & \cos \omega_1 & - {\mathfrak{C}} \cos \omega_1 & 0 \\
- {\mathfrak{C}} \sin \omega_2 & 0 & \cos \omega_2 & \sin \omega_2 \\ 
- {\mathfrak{C}} \cos \omega_2 & 0 & - \sin \omega_2 & \cos \omega_2
\end{pmatrix}. \label{MatrixG}
\end{equation}
Consider now a linear canonical transformation specified by a symplectic matrix ${\widehat{\mathbf{R}}}$ 
\begin{equation}
{\mathbfcal{Z}}_n = {\widehat{\mathbf{R}}} {\mathbfcal{Z}}_n^{\prime}, \label{LinSymplTran}
\end{equation}
which converts the matrix ${\widehat{\mathbf{G}}}$ in a block-diagonal form. Since 
\begin{equation}
{\mathbfcal{Z}}_{n+1}^{\prime} = {\widehat{\mathbf{R}}}^{-1} {\mathbfcal{Z}}_{n+1} = {\widehat{\mathbf{R}}}^{-1} {\widehat{\mathbf{G}}} {\mathbfcal{Z}}_n = {\widehat{\mathbf{R}}}^{-1} {\widehat{\mathbf{G}}} {\widehat{\mathbf{R}}} {\mathbfcal{Z}}_n^{\prime} = {\widehat{\mathbf{A}}} {\mathbfcal{Z}}_n^{\prime}, \nonumber
\end{equation}
by requirement the new symplectic matrix ${\widehat{\mathbf{A}}} = {\widehat{\mathbf{R}}}^{-1} {\widehat{\mathbf{G}}} {\widehat{\mathbf{R}}}$ should be block-diagonal 
\begin{equation}
{\widehat{\mathbf{A}}} = 
\begin{pmatrix}
{\widehat{\mathbf{A}}}_1 & {\widehat{\mathbf{0}}} \\
{\widehat{\mathbf{0}}} & {\widehat{\mathbf{A}}}_2
\end{pmatrix}. \label{MatrixA}
\end{equation}
Here ${\widehat{\mathbf{A}}}_{1,2}$ are yet unknown $2 \times 2$ matrices, and ${\widehat{\mathbf{0}}}$ is the $2 \times 2$ null matrix. Thus the basic equation to be analysed in what follows can be written as 
\begin{equation}
{\widehat{\mathbf{G}}} = {\widehat{\mathbf{R}}} {\widehat{\mathbf{A}}} {\widehat{\mathbf{R}}}^{-1}. \label{BasicEquat}
\end{equation}

Similar to the matrix ${\widehat{\mathbf{A}}}$, it is convenient to write the matrices ${\widehat{\mathbf{G}}}$ and ${\widehat{\mathbf{R}}}$ in a $2 \times 2$-block form 
\begin{equation}
{\widehat{\mathbf{G}}} = 
\begin{pmatrix}
{\widehat{\mathbf{G}}}_1 & {\widehat{\mathbf{g}}}_2 \\
{\widehat{\mathbf{g}}}_1 & {\widehat{\mathbf{G}}}_2
\end{pmatrix} \qquad {\widehat{\mathbf{R}}} = 
\begin{pmatrix}
{\widehat{\mathbf{R}}}_1 & {\widehat{\mathbf{r}}}_2 \\
{\widehat{\mathbf{r}}}_1 & {\widehat{\mathbf{R}}}_2
\end{pmatrix}, \label{MatrixGandR}
\end{equation}
and rewrite Eq. (\ref{BasicEquat}) in explicit form 
\begin{equation}
\begin{pmatrix}
{\widehat{\mathbf{G}}}_1 & {\widehat{\mathbf{g}}}_2 \\
{\widehat{\mathbf{g}}}_1 & {\widehat{\mathbf{G}}}_2
\end{pmatrix} = 
\begin{pmatrix}
{\widehat{\mathbf{R}}}_1 & {\widehat{\mathbf{r}}}_2 \\
{\widehat{\mathbf{r}}}_1 & {\widehat{\mathbf{R}}}_2
\end{pmatrix} \begin{pmatrix}
{\widehat{\mathbf{A}}}_1 & {\widehat{\mathbf{0}}} \\
{\widehat{\mathbf{0}}} & {\widehat{\mathbf{A}}}_2
\end{pmatrix} \begin{pmatrix}
{\widehat{\mathbf{R}}}_1^c & {\widehat{\mathbf{r}}}_1^c \\
{\widehat{\mathbf{r}}}_2^c & {\widehat{\mathbf{R}}}_2^c
\end{pmatrix}. \label{ExplGenRel}
\end{equation}
Here ${\widehat{\mathbf{Z}}}^c$ denotes the symplectic conjugate of the generic matrix ${\widehat{\mathbf{Z}}}$ defined in Appendix \ref{app:sympmat}. In addition, the property (\ref{ImpSympConj}) of symplectic matrices has been used to explicitly represent the above equation. 

It can be shown that the stability properties of the matrices ${\widehat{\mathbf{A}}}_{1,2}$ depend only on the matrix elements of the linear coupling matrix ${\widehat{\mathbf{G}}}$ and are independent of the particular form chosen for the matrices ${\widehat{\mathbf{R}}}_{1,2}$. Details of the derivation of their explicit form 
\begin{equation}
{\widehat{\mathbf{A}}}_1 = {\widehat{\mathbf{R}}}_1^{-1} {\left[ {\widehat{\mathbf{G}}}_1 + {\frac {1} {U D}} {\widehat{\mathbf{g}}}_2 {\left( {\widehat{\mathbf{g}}}_1 + {\widehat{\mathbf{g}}}_2^c \right)} \right]} {\widehat{\mathbf{R}}}_1, \label{MatrixA1}
\end{equation}
\begin{equation}
{\widehat{\mathbf{A}}}_2 = {\widehat{\mathbf{R}}}_2^{-1} {\left[ {\widehat{\mathbf{G}}}_2 - {\frac {1} {U D}} {\left( {\widehat{\mathbf{g}}}_1 + {\widehat{\mathbf{g}}}_2^c \right)} {\widehat{\mathbf{g}}}_2 \right]} {\widehat{\mathbf{R}}}_2, \label{MatrixA2}
\end{equation}
can be found in Appendix \ref{app:matrix12}. The quantities $U$ and $D$ entering the right-hand-sides of the above equations are given according to the expressions in Eqs. (\ref{SymplRProp}) -- (\ref{DefConst4}). An important comment is now in order. As it is known, the dynamic properties described by the roots of the characteristic polynomial of a $2 \times 2$ symplectic matrix are characterized solely by its trace. Since the similarity transformation leaves the trace of a generic matrix invariant, it follows that the stability of motion depends only on the matrices in the square brackets in the expressions above, which are expressed solely by the elements of the matrix ${\widehat{\mathbf{G}}}$. This means that there is some freedom in the choice of matrices ${\widehat{\mathbf{R}}}_{1,2}$.

The relevant quantities for the specific case considered here can be expressed as 
\begin{eqnarray}
T = 2 {\left( \cos \omega_1 - \cos \omega_2 \right)}, \nonumber 
\\
2 {\left( \cos \omega_1 + \cos \omega_2 \right)} = G_1 +G_2 \nonumber 
\\
= A_1 + A_2 = 2 {\left( \cos \Omega_1 + \cos \Omega_2 \right)}. \label{ConsTplusProp}
\end{eqnarray}
Furthermore, 
\begin{eqnarray}
D && = {\frac {1} {2}} {\left[ 1 + {\frac {\cos \omega_1 - \cos \omega_2} {\cos \Omega_1 - \cos \Omega_2}} \sgn {\left( \cos \omega_1 - \cos \omega_2 \right)} \right]}, \nonumber 
\\ 
&& U = 2 {\left( \cos \Omega_1 - \cos \Omega_2 \right)} \sgn {\left( \cos \omega_1 - \cos \omega_2 \right)}, 
\label{ConstDandU}
\end{eqnarray}
\begin{eqnarray}
{\mathfrak{C}} \sin \omega_1 && \sin \omega_2 = {\left( \cos \Omega_1 - \cos \Omega_2 + \cos \omega_1 - \cos \omega_2 \right)} \nonumber 
\\ 
&& \times {\left( \cos \Omega_1 - \cos \Omega_2 - \cos \omega_1 + \cos \omega_2 \right)}, \label{PerturbC}
\end{eqnarray}
where $\sgn{\left( x \right)}$ denotes the sign of the corresponding variable $x$. According to the mentioned above, we have a certain freedom in the choice of the matrices ${\widehat{\mathbf{R}}}_{1,2}$, so that the simplest choice consists in the convention that they are proportional to the unity matrix
\begin{equation}
{\widehat{\mathbf{R}}}_{1,2} = {\sqrt{D}} {\widehat{\mathbf{I}}}. \label{MatrixR12}
\end{equation}
Consider first the case, where $\sgn{\left( \cos \omega_1 - \cos \omega_2 \right)} = 1$. Obviously Eqs. (\ref{MatrixA1}) and (\ref{MatrixA2}) can be rewritten as 
\begin{widetext}
\begin{eqnarray}
{\widehat{\mathbf{A}}}_1 = {\widehat{\mathbf{G}}}_1 + {\frac {1} {U D}} {\widehat{\mathbf{g}}}_2 {\left( {\widehat{\mathbf{g}}}_1 + {\widehat{\mathbf{g}}}_2^c \right)} = 
\begin{pmatrix}
\cos \omega_2 + \cos \Omega_1 - \cos \Omega_2 & \sin \omega_1 \\
- \sin \omega_1 + {\left( \cos \Omega_1 -\cos \Omega_2 - \cos \omega_1 + \cos \omega_2 \right)} \cot \omega_1 & \cos \omega_1
\end{pmatrix}, \nonumber 
\\ 
{\widehat{\mathbf{A}}}_2 = {\widehat{\mathbf{G}}}_2 - {\frac {1} {U D}} {\left( {\widehat{\mathbf{g}}}_1 + {\widehat{\mathbf{g}}}_2^c \right)} {\widehat{\mathbf{g}}}_2 = 
\begin{pmatrix}
\cos \omega_1 - \cos \Omega_1 + \cos \Omega_2 & \sin \omega_2 \\
- \sin \omega_2 - {\left( \cos \Omega_1 -\cos \Omega_2 - \cos \omega_1 + \cos \omega_2 \right)} \cot \omega_2 & \cos \omega_2
\end{pmatrix}, \label{MatrixA12}
\end{eqnarray}
\end{widetext} 
If $\sgn{\left( \cos \omega_1 - \cos \omega_2 \right)} = -1$, we obtain expressions for the normal form matrices ${\widehat{\mathbf{A}}}_1$ and ${\widehat{\mathbf{A}}}_2$ similar to the above ones, but with $\Omega_1$ and $\Omega_2$ interchanged. In what follows, we shall consider in detail the case where $\sgn{\left( \cos \omega_1 - \cos \omega_2 \right)} = 1$ - the opposite sign case can be treated in analogous way. The other two blocks of the transformation matrix ${\widehat{\mathbf{R}}}$ can be determined according to the chain of expressions in Eq. (\ref{Equatg1+g2c}). Thus, we have 
\begin{eqnarray}
{\widehat{\mathbf{r}}}_1 = {\frac {{\widehat{\mathbf{g}}}_1 + {\widehat{\mathbf{g}}}_2^c} {U {\sqrt{D}}}} = {\frac {{\mathfrak{C}}} {U {\sqrt{D}}}}
\begin{pmatrix}
- \sin \omega_2 & 0 \\
T / 2 & - \sin \omega_1
\end{pmatrix} \nonumber 
\\
{\widehat{\mathbf{r}}}_2 = - {\widehat{\mathbf{r}}}_1^c = {\frac {{\mathfrak{C}}} {U {\sqrt{D}}}}
\begin{pmatrix}
\sin \omega_1 & 0 \\
T / 2 & \sin \omega_2
\end{pmatrix}. \label{MatrixSmR12}
\end{eqnarray}

From the normal form matrices ${\widehat{\mathbf{A}}}_1$ and ${\widehat{\mathbf{A}}}_2$ given explicitly by Eq. (\ref{MatrixA12}), the normal mode Twiss parameters $\alpha_i$, $\beta_i$ and $\gamma_i$ for $i = 1, 2$ can be determined using the standard expression for the one-turn transfer matrix \cite{TzenovBOOK}  
\begin{equation}
{\widehat{\mathbf{A}}}_i = 
\begin{pmatrix}
\cos \Omega_i + \alpha_i \sin \Omega_i & \beta_i \sin \Omega_i \\
- \gamma_i \sin \Omega_i & \cos \Omega_i - \alpha_i \sin \Omega_i
\end{pmatrix}. \label{TwissMatrix}
\end{equation}
The result is 
\begin{equation}
\alpha_1 = {\frac {\cos \Omega_1 - \cos \Omega_2 - \cos \omega_1 + \cos \omega_2} {2 \sin \Omega_1}}, \quad \beta_1 = {\frac {\sin \omega_1} {\sin \Omega_1}}, \label{Twiss11}
\end{equation}
\begin{equation}
\gamma_1 = \beta_1 - {\frac {\cot \omega_1} {\sin \Omega_1}} {\left( \cos \Omega_1 - \cos \Omega_2 - \cos \omega_1 + \cos \omega_2 \right)}, \label{Twiss12}
\end{equation}
\begin{equation}
\alpha_2 = {\frac {\cos \Omega_2 - \cos \Omega_1 + \cos \omega_1 - \cos \omega_2} {2 \sin \Omega_2}}, \quad \beta_2 = {\frac {\sin \omega_2} {\sin \Omega_2}}, \label{Twiss21}
\end{equation}
\begin{equation}
\gamma_2 = \beta_2 - {\frac {\cot \omega_2} {\sin \Omega_2}} {\left( \cos \Omega_2 - \cos \Omega_1 + \cos \omega_1 - \cos \omega_2 \right)}, \label{Twiss22}
\end{equation}
where by direct substitution it can be verified that $\beta_i \gamma_i - \alpha_i^2 = 1$ for $i = 1, 2$, as should be expected. Since the motion in the normal mode is decoupled, there exist the two independent Courant-Snyder invariants 
\begin{equation}
I_n^{(1)} = \gamma_1 X_n^{\prime 2} + 2 \alpha_1 X_n^{\prime} P_{x,n}^{\prime} + \beta_1 P_{x,n}^{\prime 2}. \label{CourSnydInv1}
\end{equation}
\begin{equation}
I_n^{(2)} = \gamma_2 Z_n^{\prime 2} + 2 \alpha_2 Z_n^{\prime} P_{z,n}^{\prime} + \beta_2 P_{z,n}^{\prime 2}. \label{CourSnydInv2}
\end{equation}

Our final task consists in expressing the above invariants in terms of the original canonical variables ${\mathbfcal{Z}}_n$. By inverting the linear canonical transformation defined in Eq. (\ref{LinSymplTran}), we can write 
\begin{equation}
{\mathbfcal{Z}}_n^{\prime} = {\widehat{\mathbf{R}}}^{-1} {\mathbfcal{Z}}_n = 
\begin{pmatrix}
{\widehat{\mathbf{R}}}_1^c & {\widehat{\mathbf{r}}}_1^c \\
{\widehat{\mathbf{r}}}_2^c & {\widehat{\mathbf{R}}}_2^c
\end{pmatrix} {\mathbfcal{Z}}_n = 
\begin{pmatrix}
{\sqrt{D}} {\widehat{\mathbf{I}}} & - {\widehat{\mathbf{r}}}_2 \\
- {\widehat{\mathbf{r}}}_1 & {\sqrt{D}} {\widehat{\mathbf{I}}}
\end{pmatrix} {\mathbfcal{Z}}_n, \label{InvSymplTran}
\end{equation}
or alternatively 
\begin{equation}
\begin{pmatrix}
X_n^{\prime} \\
P_{x,n}^{\prime} 
\end{pmatrix} = 
\begin{pmatrix}
{\sqrt{D}} X_n - {\dfrac {{\mathfrak{C}}} {U {\sqrt{D}}}} Z_n \sin \omega_1 \\
{\sqrt{D}} P_{x,n} - {\dfrac {{\mathfrak{C}}} {U {\sqrt{D}}}} {\left( {\dfrac {T} {2}} Z_n + P_{z,n} \sin \omega_2 \right)}
\end{pmatrix}, \label{XStatVec}
\end{equation}
\begin{equation}
\begin{pmatrix}
Z_n^{\prime} \\
P_{z,n}^{\prime} 
\end{pmatrix} = 
\begin{pmatrix}
{\sqrt{D}} Z_n + {\dfrac {{\mathfrak{C}}} {U {\sqrt{D}}}} X_n \sin \omega_2 \\
{\sqrt{D}} P_{z,n} - {\dfrac {{\mathfrak{C}}} {U {\sqrt{D}}}} {\left( {\dfrac {T} {2}} X_n - P_{x,n} \sin \omega_1 \right)}
\end{pmatrix}. \label{ZStatVec}
\end{equation}
\begin{figure}[!htb]
\centering
\includegraphics[width=\hsize]{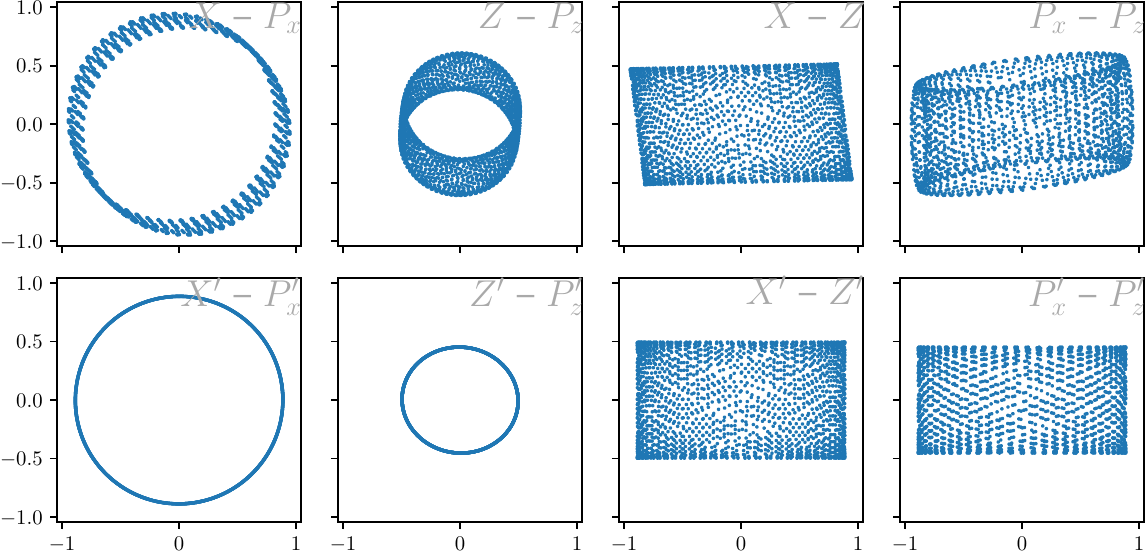}
\caption{The linear map (\ref{CoupledMapX}) -- (\ref{CoupledMapPZ}) has been iterated 2000 turns with respect to the following randomly chosen (sufficiently far from resonances) parameters: $\nu_1=0.75$, $\nu_2=0.53$, and $\mathfrak{C}=0.25$. The initial state vector has been set to ${\mathbfcal Z}_0 = [0.3,0.8,-0.3,0.5]^T$ but it is almost irrelevant for the shape of the trajectories in phase space. First row: phase space projections in the initial coordinates ${\mathbfcal Z}$; second row: phase space projections in the normal form coordinates ${\mathbfcal Z}^{\prime}$. First and second column: horizontal, respectively vertical phase space projections; third column: transverse plane trajectory.}
\label{fig:sim}
\end{figure}

What remains to be done now is to replace the new phase space coordinates ${\mathbfcal{Z}}_n^{\prime}$ with the corresponding expressions in terms of the initial ones ${\mathbfcal{Z}}_n$ given by the above equations, in the Courant-Snyder invariants defined by Eqs. (\ref{CourSnydInv1}) and (\ref{CourSnydInv2}). Thus, we obtain the sought-for two independent invariants in the initial coordinates in phase space. And so, our original goal has been reached; a split description of the coupled betatron motion in terms of new optical eigenfunctions (normal mode Twiss parameters) defined in a new coordinate system has been found.

\section{\label{sec:simulation}Tracking and Illustration of the Beam Dynamics}

The one-turn map given by Eqs. (\ref{CoupledMapX}) -- (\ref{CoupledMapPZ}) and describing the linear betatron coupling was iterated 2000 turns for different values of the coupling coefficient $\mathfrak{C}$. Unfortunately, it is not possible to visualize the multidimensional torus (on which the phase-space trajectory lies) in the full four-dimensional phase space. For this reason, the initial state vector ${\mathbfcal Z}_0$ has been evolved and after each turn has been mapped as a point on the corresponding sub-spaces of the full four-dimensional phase space. 

Focusing a look at the simulation results (as shown in Figure \ref{fig:sim}), one can observe that particle evolution in the four-dimensional phase space is actually a trajectory on a higher-dimensional (four-dimensional) torus spanned over the horizontal and the vertical two-dimensional phase spaces. 

The inverse canonical transformation specified by the matrix ${\widehat{\mathbf{R}}}^{-1}$ and given by Eqs. (\ref{XStatVec}) and (\ref{ZStatVec}) reveals the major orbits in the phase space. By keeping the unperturbed betatron tunes $\nu_x$ and $\nu_z$ unchanged and adjusting the coupling coefficient $\mathfrak{C}$, which relates the disposition of solenoids and the skew quadruples along the ring circumference, one can follow in detail the metamorphosis of shape change of the phase-space manifold from a ring torus to a horn torus and then a spindle torus. It is worth also to note that the trajectory on the manifold exhibits some fine structure determined by the map. At first glance, it may seem that trajectories with the same initial state vector lie on the same manifold and intersect with each other. What is shown in the figure, however, is a projection onto the horizontal/vertical plane in the phase space, so that it does not violate the Liouville theorem.

Mentioned last but not least important, at each iteration step (that is, after each turn) a check has been carried out, which showed that the invariants (\ref{CourSnydInv1}) and (\ref{CourSnydInv2}) are preserved.

\section{\label{sec:conclrem}Concluding Remarks} 

In all cases of practical interest numerical methods for solving the differential equations governing particle motion involve discretization schemes anyway. The substitution of the Hamilton’s equations of motion with mappings is a natural way to alternatively describe particle dynamics. Using the technique of the discrete one-turn transfer maps, the problem of linear coupling between horizontal and vertical betatron oscillations in an accelerator has been treated exactly and entirely in explicit form.

The stability region of betatron oscillations with linear coupling between the transverse degrees of freedom in the fractional part of the horizontal and the vertical betatron tune space as a function of the linear coupling strength, has been obtained. As far as our knowledge of the matter extends, this result is being reported for the first time. It is intuitively clear to expect the instability regions to be located in the vicinity of the linear sum resonances of the form $\Frac {\left( \nu_x \right)} + \Frac {\left( \nu_z \right)} = 0$ and $\Frac {\left( \nu_x \right)} + \Frac {\left( \nu_z \right)} = \pm 1$, where $\Frac {\left( \nu_{x,z} \right)}$ are the fractional parts of the betatron tunes. It has been also shown that the increment/decrement of the horizontal and the vertical betatron oscillations in the case of the linear sum resonance is approximately equal to the half of the coupling strength. 

Further, the normal form parameterization of the one-turn linear map is worked out in detail. It has been shown that the normal form representation possesses an important feature that the stability properties of both the $2 \times 2$ symplectic matrices comprising the diagonal of the block-diagonal transfer matrix in the normal form depend only on the matrix elements of the original linear coupling matrix and are independent of the particular form chosen for the diagonal $2 \times 2$ matrix blocks of the symplectic transformation matrix bringing the initial one-turn matrix to normal form. Since by construction the motion in the normal mode in the new normal form coordinates is decoupled, there must exist two independent Courant-Snyder invariants, which have been found explicitly.

The systematic developments presented here provide a normal form parameterization for the four-dimensional symplectic one-turn matrix, which has a close connection to the original Courant-Snyder representation of the two-dimensional symplectic
matrix. All of the parameters and expressions entering explicitly the transfer map parameterization can provide a valuable framework for accelerator design and particle simulation studies.

\begin{acknowledgments}

Fruitful discussions with Drs. Bart Faatz and Jianhui Chen are gratefully acknowledged. We would like to thank Prof. Zhentang Zhao for his interest in the present work, and for his support.

One of us (SIT) wishes to acknowledge the support from the Alliance of International Science Organizations Grant No. ANSO-VF-2021-05.

\end{acknowledgments}

\appendix

\section{\label{app:boundaries}Boundaries of the Stability Diagram}

The stability constrain~(Eq.~\eqref{Stability}) can be split into the following inequalities:
\begin{widetext}
\begin{align}
\cos\omega_1 + \cos\omega_2 + \sqrt{(\cos\omega_1 - \cos\omega_2)^2 +\mathfrak{C}^2 \sin\omega_1 \sin\omega_2} &\leqslant 2, \\
\cos\omega_1 + \cos\omega_2 - \sqrt{(\cos\omega_1 - \cos\omega_2)^2 +\mathfrak{C}^2 \sin\omega_1 \sin\omega_2} &\geqslant -2, \\
(\cos\omega_1 - \cos\omega_2)^2 +\mathfrak{C}^2 \sin\omega_1 \sin\omega_2 &\geqslant 0.
\end{align}
\end{widetext}

Obviously, there exists the trivial solutions:
\begin{align}
\sin\omega_1\sin\omega_2 &=0 \\
\intertext{i.e.,}
\omega_1 &= k_1\pi, \\
\omega_2 &= k_2\pi, 
\end{align}
where $k_1,k_2 \in \mathbb{Z}$.

\begin{figure}[!htb]
    \centering
    \includegraphics[width=\hsize]{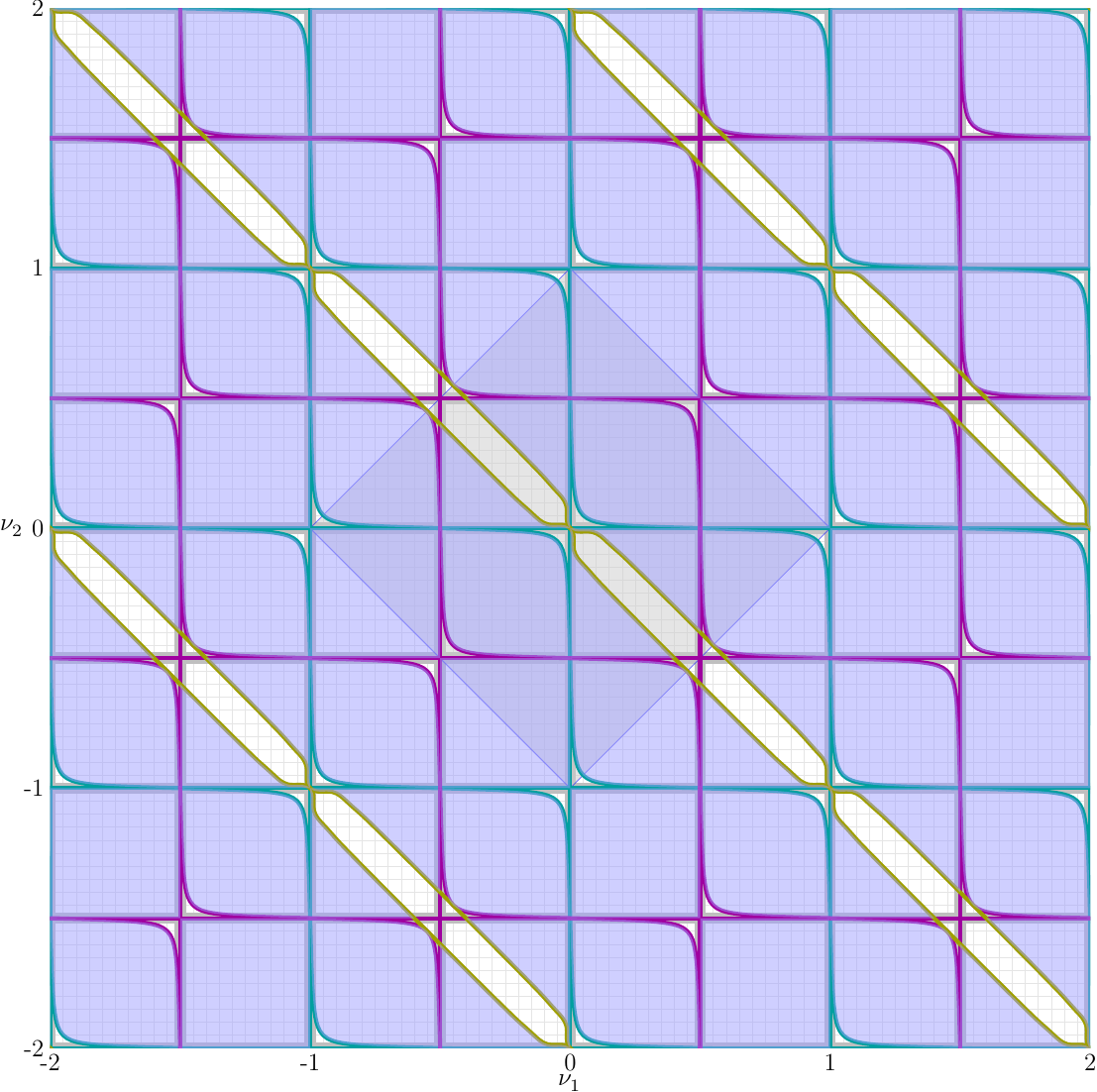}
    \caption{Stability areas (with $\mathfrak{C}^2=0.1$) divided by the constrains. The cyan curves are the solutions~\eqref{eq:stability:sol2} and the trivial solutions~\eqref{eq:stability:sol4}; the magenta curves are the solutions~\eqref{eq:stability:sol1} and the trivial solutions~\eqref{eq:stability:sol5}; and the yellow curves are the solutions~\eqref{eq:stability:sol3}. The shaded square at the center is one of the tiles.}
    \label{fig:stability-diagram}
\end{figure}

After transforming the coordinates with the following rule
\begin{equation}
    \begin{pmatrix}
    u\\v
    \end{pmatrix}
    = \frac12
    \begin{pmatrix}
    1 & -1\\
    1 & 1
    \end{pmatrix}
    \begin{pmatrix}
    \omega_1 \\ \omega_2
    \end{pmatrix} ,
\end{equation}
one can obtain a set of simple boundaries:
\begin{align}
\cos v &= \pm \frac{\mathfrak{C}^2 - 4}{\mathfrak{C}^2 + 4} \cos u, \\
\cos v &= \frac{2\cos u - \mathfrak{C}^2 \cos u - 2}{2\cos u - \mathfrak{C}^2 - 2}.
\end{align}
The analytical boundaries of the stability areas are straightforward (as shown in Fig.~\ref{fig:stability-diagram}):
\begin{align}
v &= \pm \arccos\left(  \frac{\mathfrak{C}^2 - 4}{\mathfrak{C}^2 + 4} \cos u \right) + 2k_3\pi, \label{eq:stability:sol1}\\
v &= \pm \arccos\left( -\frac{\mathfrak{C}^2 - 4}{\mathfrak{C}^2 + 4} \cos u \right) + 2k_4\pi, \label{eq:stability:sol2}\\
v &= \pm \arccos\left( \frac{2\cos u - \mathfrak{C}^2 \cos u - 2}{2\cos u - \mathfrak{C}^2 - 2} \right) + 2k_5\pi, \label{eq:stability:sol3}
\end{align}
where $k_3,k_4,k_5 \in \mathbb{Z}$. We can also rewrite the trivial solutions:
\begin{align}
v &= \pm u + 2k_6\pi, \label{eq:stability:sol4}\\
v &= \pm u + (2k_7-1)\pi, \label{eq:stability:sol5}
\end{align}
and again $k_6,k_7 \in \mathbb{Z}$.

\section{\label{app:sympmat}Review of Some Basic Properties of Symplectic Matrices}

By definition, the four-by-four matrix ${\widehat{\mathbf{S}}}$ is symplectic if
\begin{equation}
{\widehat{\mathbf{S}}}^T {\widehat{\mathbfcal{J}}} {\widehat{\mathbf{S}}} = {\widehat{\mathbfcal{J}}}, \quad \text{where } {\widehat{\mathbfcal{J}}} = 
\begin{pmatrix}
{\widehat{\mathbf{J}}} & {\widehat{\mathbf{0}}} \\
{\widehat{\mathbf{0}}} & {\widehat{\mathbf{J}}}
\end{pmatrix}, \quad {\widehat{\mathbf{J}}} = 
\begin{pmatrix}
0 & 1 \\
-1 & 0
\end{pmatrix}, \label{DefSympMat}
\end{equation}
and the superscript ``$T$'' implies matrix transposition. The basic nonsingular, skew-symmetric matrix ${\widehat{\mathbfcal{J}}}$ has the obvious properties 
\begin{equation}
{\widehat{\mathbfcal{J}}}^2 = - {\widehat{\mathbf{I}}}, \qquad {\widehat{\mathbfcal{J}}}^{-1} = - {\widehat{\mathbfcal{J}}}, \qquad {\widehat{\mathbfcal{J}}}^T = - {\widehat{\mathbfcal{J}}}. \label{PropSkewMat}
\end{equation}
From the equation 
\begin{equation}
{\widehat{\mathbf{S}}} {\widehat{\mathbfcal{J}}} {\widehat{\mathbf{S}}}^T {\widehat{\mathbfcal{J}}} {\widehat{\mathbf{S}}} = {\widehat{\mathbf{S}}} {\widehat{\mathbfcal{J}}}^2 = - {\widehat{\mathbf{S}}} {\widehat{\mathbf{I}}} = - {\widehat{\mathbf{I}}} {\widehat{\mathbf{S}}}, \nonumber
\end{equation}
an alternative definition 
\begin{equation}
{\widehat{\mathbf{S}}} {\widehat{\mathbfcal{J}}} {\widehat{\mathbf{S}}}^T = {\widehat{\mathbfcal{J}}}, \label{AltDefSympMat}
\end{equation}
of a symplectic matrix follows. Next, we define the symplectic conjugate \cite{Courant} of a generic matrix ${\widehat{\mathbfcal{A}}}$ to be 
\begin{equation}
{\widehat{\mathbfcal{A}}}^c = - {\widehat{\mathbfcal{J}}} {\widehat{\mathbfcal{A}}}^T {\widehat{\mathbfcal{J}}}. \label{SympConj}
\end{equation}
From Eq. (\ref{DefSympMat}) or Eq. (\ref{AltDefSympMat}) an important property 
\begin{equation}
{\widehat{\mathbf{S}}}^c = {\widehat{\mathbf{S}}}^{-1}, \label{ImpSympConj}
\end{equation}
of symplectic matrices follows. For a generic $2 \times 2$ matrix ${\widehat{\mathbf{B}}}$, we have 
\begin{equation}
{\widehat{\mathbf{B}}} = 
\begin{pmatrix}
b_{11} & b_{12} \\
b_{21} & b_{22}
\end{pmatrix}, \qquad \qquad {\widehat{\mathbf{B}}}^c = 
\begin{pmatrix}
b_{22} & - b_{12} \\
- b_{21} & b_{11}
\end{pmatrix}, \label{Gen2By2Prop1}
\end{equation}
and 
\begin{eqnarray}
{\widehat{\mathbf{B}}} {\widehat{\mathbf{B}}}^c = {\widehat{\mathbf{B}}}^c {\widehat{\mathbf{B}}} = {\widehat{\mathbf{I}}} \det{\left( {\widehat{\mathbf{B}}} \right)}, \nonumber
\\ 
{\widehat{\mathbf{B}}} + {\widehat{\mathbf{B}}}^c = {\widehat{\mathbf{I}}} \Sp{\left( {\widehat{\mathbf{B}}} \right)}, \nonumber
\\ 
{\widehat{\mathbf{B}}}^c = {\widehat{\mathbf{B}}}^{-1} \det{\left( {\widehat{\mathbf{B}}} \right)}, \label{Gen2By2Prop2}
\end{eqnarray}

where as usual ``$\det$'' and ``$\Sp$'' denote the determinant and the trace of the dedicated matrix, respectively.

Let us write the $4 \times 4$ symplectic matrix ${\widehat{\mathbf{S}}}$ in a $2 \times 2$ block form 
\begin{equation}
{\widehat{\mathbf{S}}} = 
\begin{pmatrix}
{\widehat{\mathbf{S}}}_1 & {\widehat{\mathbf{s}}}_2 \\
{\widehat{\mathbf{s}}}_1 & {\widehat{\mathbf{S}}}_2
\end{pmatrix}, \label{BlockS}
\end{equation}
Since the inverse of a symplectic matrix is equal to its symplectic conjugate, according to Eq. (\ref{ImpSympConj}), we must have ${\widehat{\mathbf{S}}} {\widehat{\mathbf{S}}}^c = {\widehat{\mathbf{S}}}^c {\widehat{\mathbf{S}}} = {\widehat{\mathbf{I}}}$. Comparing 
\begin{eqnarray}
{\widehat{\mathbf{S}}} {\widehat{\mathbf{S}}}^c = 
\begin{pmatrix}
{\widehat{\mathbf{S}}}_1 & {\widehat{\mathbf{s}}}_2 \\
{\widehat{\mathbf{s}}}_1 & {\widehat{\mathbf{S}}}_2
\end{pmatrix} \begin{pmatrix}
{\widehat{\mathbf{S}}}^c_1 & {\widehat{\mathbf{s}}}^c_1 \\
{\widehat{\mathbf{s}}}^c_2 & {\widehat{\mathbf{S}}}^c_2
\end{pmatrix} \nonumber 
\\ 
= \begin{pmatrix}
{\widehat{\mathbf{S}}}_1 {\widehat{\mathbf{S}}}_1^c + {\widehat{\mathbf{s}}}_2 {\widehat{\mathbf{s}}}_2^c & {\widehat{\mathbf{S}}}_1 {\widehat{\mathbf{s}}}^c_1 + {\widehat{\mathbf{s}}}_2 {\widehat{\mathbf{S}}}^c_2 \\
{\widehat{\mathbf{s}}}_1 {\widehat{\mathbf{S}}}_1^c + {\widehat{\mathbf{S}}}_2 {\widehat{\mathbf{s}}}^c_2 & {\widehat{\mathbf{S}}}_2 {\widehat{\mathbf{S}}}_2^c + {\widehat{\mathbf{s}}}_1 {\widehat{\mathbf{s}}}_1^c
\end{pmatrix}, \nonumber
\end{eqnarray}
and 
\begin{eqnarray}
{\widehat{\mathbf{S}}}^c {\widehat{\mathbf{S}}} = 
\begin{pmatrix}
{\widehat{\mathbf{S}}}_1^c & {\widehat{\mathbf{s}}}_1^c \\
{\widehat{\mathbf{s}}}_2^c & {\widehat{\mathbf{S}}}_2^c
\end{pmatrix} \begin{pmatrix}
{\widehat{\mathbf{S}}}_1 & {\widehat{\mathbf{s}}}_2 \\
{\widehat{\mathbf{s}}}_1 & {\widehat{\mathbf{S}}}_2
\end{pmatrix} \nonumber 
\\ 
= \begin{pmatrix}
{\widehat{\mathbf{S}}}_1^c {\widehat{\mathbf{S}}}_1 + {\widehat{\mathbf{s}}}_1^c {\widehat{\mathbf{s}}}_1 & {\widehat{\mathbf{S}}}_1^c {\widehat{\mathbf{s}}}_2 + {\widehat{\mathbf{s}}}_1^c {\widehat{\mathbf{S}}}_2 \\
{\widehat{\mathbf{s}}}_2^c {\widehat{\mathbf{S}}}_1 + {\widehat{\mathbf{S}}}_2^c {\widehat{\mathbf{s}}}_1 & {\widehat{\mathbf{S}}}_2^c {\widehat{\mathbf{S}}}_2 + {\widehat{\mathbf{s}}}_2^c {\widehat{\mathbf{s}}}_2
\end{pmatrix}, \nonumber
\end{eqnarray}
by using Eq. (\ref{Gen2By2Prop2}), we obtain 
\begin{eqnarray}
\det{\left( {\widehat{\mathbf{S}}}_1 \right)} + \det{\left( {\widehat{\mathbf{s}}}_2 \right)} && = 1, \qquad \det{\left( {\widehat{\mathbf{S}}}_2 \right)} + \det{\left( {\widehat{\mathbf{s}}}_1 \right)} = 1, \nonumber 
\\ 
&& {\widehat{\mathbf{S}}}_1 {\widehat{\mathbf{s}}}^c_1 + {\widehat{\mathbf{s}}}_2 {\widehat{\mathbf{S}}}^c_2 = 0, \label{SymplSProp1}
\end{eqnarray}
and 
\begin{eqnarray}
\det{\left( {\widehat{\mathbf{S}}}_1 \right)} + \det{\left( {\widehat{\mathbf{s}}}_1 \right)} && = 1, \qquad \det{\left( {\widehat{\mathbf{S}}}_2 \right)} + \det{\left( {\widehat{\mathbf{s}}}_2 \right)} = 1, \nonumber 
\\ 
&& {\widehat{\mathbf{S}}}_1^c {\widehat{\mathbf{s}}}_2 + {\widehat{\mathbf{s}}}_1^c {\widehat{\mathbf{S}}}_2 = 0, \label{SymplSProp2}
\end{eqnarray}
respectively. Note that the above relations also imply the following properties 
\begin{equation}
\det{\left( {\widehat{\mathbf{S}}}_1 \right)} = \det{\left( {\widehat{\mathbf{S}}}_2 \right)}, \qquad \det{\left( {\widehat{\mathbf{s}}}_1 \right)} = \det{\left( {\widehat{\mathbf{s}}}_2 \right)}. \label{SymplSProp3}
\end{equation}
Equations (\ref{SymplSProp1}) and (\ref{SymplSProp2}) are actually equivalent, and they impose a total of 6 independent constraints on the 16 matrix elements of ${\widehat{\mathbf{S}}}$. The four-by-four symplectic matrix ${\widehat{\mathbf{S}}}$, is
therefore specified by 10 independent parameters. 

\section{\label{app:matrix12}Derivation of Eqs. (\ref{MatrixA1}) and (\ref{MatrixA2})}

Carrying out explicit the matrix multiplications in Eq. (\ref{ExplGenRel}), we find
\begin{equation}
{\widehat{\mathbf{G}}}_1 = {\widehat{\mathbf{R}}}_1 {\widehat{\mathbf{A}}}_1 {\widehat{\mathbf{R}}}_1^c + {\widehat{\mathbf{r}}}_2 {\widehat{\mathbf{A}}}_2 {\widehat{\mathbf{r}}}_2^c \qquad {\widehat{\mathbf{G}}}_2 = {\widehat{\mathbf{R}}}_2 {\widehat{\mathbf{A}}}_2 {\widehat{\mathbf{R}}}_2^c + {\widehat{\mathbf{r}}}_1 {\widehat{\mathbf{A}}}_1 {\widehat{\mathbf{r}}}_1^c, \label{MatrixG12}
\end{equation}
\begin{equation}
{\widehat{\mathbf{g}}}_1 = {\widehat{\mathbf{r}}}_1 {\widehat{\mathbf{A}}}_1 {\widehat{\mathbf{R}}}_1^c + {\widehat{\mathbf{R}}}_2 {\widehat{\mathbf{A}}}_2 {\widehat{\mathbf{r}}}_2^c \qquad {\widehat{\mathbf{g}}}_2 = {\widehat{\mathbf{R}}}_1 {\widehat{\mathbf{A}}}_1 {\widehat{\mathbf{r}}}_1^c + {\widehat{\mathbf{r}}}_2 {\widehat{\mathbf{A}}}_2 {\widehat{\mathbf{R}}}_2^c, \label{MatrixGg12}
\end{equation}
First of all, let us note that since the transformation matrix ${\widehat{\mathbf{R}}}$ is symplectic, relations similar to (\ref{SymplSProp1}) -- (\ref{SymplSProp3}) for the corresponding blocks of ${\widehat{\mathbf{R}}}$ must hold. In particular 
\begin{equation}
\det{\left( {\widehat{\mathbf{R}}}_1 \right)} = \det{\left( {\widehat{\mathbf{R}}}_2 \right)} = D, \quad \det{\left( {\widehat{\mathbf{r}}}_1 \right)} = \det{\left( {\widehat{\mathbf{r}}}_2 \right)} = 1 - D. \label{SymplRProp}
\end{equation}
Let us now define 
\begin{eqnarray}
A_1 = \Sp{\left( {\widehat{\mathbf{A}}}_1 \right)}, \qquad A_2 = \Sp{\left( {\widehat{\mathbf{A}}}_2 \right)}, \label{DefConst1}
\\ 
G_1 = \Sp{\left( {\widehat{\mathbf{G}}}_1 \right)}, \qquad G_2 = \Sp{\left( {\widehat{\mathbf{G}}}_2 \right)}, \label{DefConst2}
\end{eqnarray}
and 
\begin{equation}
T = \Sp{\left( {\widehat{\mathbf{G}}}_1 - {\widehat{\mathbf{G}}}_2 \right)} = G_1 - G_2, \label{DefConst3}
\end{equation}
\begin{equation}
U = \Sp{\left( {\widehat{\mathbf{A}}}_1 - {\widehat{\mathbf{A}}}_2 \right)} = A_1 - A_2. \label{DefConst4}
\end{equation}

Taking the trace of equations (\ref{MatrixG12}), and using (\ref{SymplRProp}), the last property in Eq. (\ref{Gen2By2Prop2}) and the invariance of the trace of a matrix subjected to a similarity transformation, we obtain 
\begin{equation}
G_1 = A_1 D + A_2 {\left( 1 - D \right)}, \quad G_2 = A_2 D + A_1 {\left( 1 - D \right)}. \label{EquatG1G2}
\end{equation}
Adding and subtracting the last two equations, important relations influencing the dynamical stability follow 
\begin{equation}
G_1 + G_2 = A_1 + A_2, \qquad T = G_1 - G_2 = U {\left( 2 D - 1 \right)}. \label{EquatG1G21}
\end{equation}
Adding the first of Eqs. (\ref{MatrixGg12}) and the symplectic conjugate of the second one, we find 
\begin{eqnarray}
{\widehat{\mathbf{g}}}_1 + {\widehat{\mathbf{g}}}_2^c && = {\widehat{\mathbf{r}}}_1 {\left( {\widehat{\mathbf{A}}}_1 + {\widehat{\mathbf{A}}}_1^c \right)} {\widehat{\mathbf{R}}}_1^c + {\widehat{\mathbf{R}}}_2 {\left( {\widehat{\mathbf{A}}}_2 + {\widehat{\mathbf{A}}}_2^c \right)} {\widehat{\mathbf{r}}}_2^c \nonumber 
\\ 
&& = A_1 {\widehat{\mathbf{r}}}_1 {\widehat{\mathbf{R}}}_1^c + A_2 {\widehat{\mathbf{R}}}_2 {\widehat{\mathbf{r}}}_2^c, \nonumber
\end{eqnarray}
where we have used the second property in Eq. (\ref{Gen2By2Prop2}). Taking into account the symplectic conjugate of the last property in Eq. (\ref{SymplSProp1}) written for the matrix blocks of ${\widehat{\mathbf{R}}}$, we finally arrive at 
\begin{equation}
{\widehat{\mathbf{g}}}_1 + {\widehat{\mathbf{g}}}_2^c = U {\widehat{\mathbf{r}}}_1 {\widehat{\mathbf{R}}}_1^c = - U {\widehat{\mathbf{R}}}_2 {\widehat{\mathbf{r}}}_2^c. \label{Equatg1+g2c}
\end{equation}
Taking now the determinant of Eq. (\ref{Equatg1+g2c}), we obtain \cite{Comment} 
\begin{equation}
\det{\left( {\widehat{\mathbf{g}}}_1 + {\widehat{\mathbf{g}}}_2^c \right)} = U^2 \det{\left( {\widehat{\mathbf{r}}}_1 \right)} \det{\left( {\widehat{\mathbf{R}}}_1 \right)} = U^2 D {\left( 1 - D \right)}. \label{Determg1+g2c}
\end{equation}
From the above equation and the second of Eqs. (\ref{EquatG1G21}), the determinant $D$ of the diagonal blocks ${\widehat{\mathbf{R}}}_{1,2}$ of the unknown transformation matrix can be expressed in terms of known quantities, namely 
\begin{equation}
{\left( 2 D - 1 \right)}^2 = {\frac {T^2} {T^2 + 4 \det{\left( {\widehat{\mathbf{g}}}_1 + {\widehat{\mathbf{g}}}_2^c \right)}}}. \label{BasDetermEq}
\end{equation}

The final step is to perform in an explicit form the following matrix multiplication
\begin{eqnarray}
{\widehat{\mathbf{g}}}_2 {\left( {\widehat{\mathbf{g}}}_1 + {\widehat{\mathbf{g}}}_2^c \right)} && = U {\left( {\widehat{\mathbf{R}}}_1 {\widehat{\mathbf{A}}}_1 {\widehat{\mathbf{r}}}_1^c {\widehat{\mathbf{r}}}_1 {\widehat{\mathbf{R}}}_1^c - 
{\widehat{\mathbf{r}}}_2 {\widehat{\mathbf{A}}}_2 {\widehat{\mathbf{R}}}_2^c {\widehat{\mathbf{R}}}_2 {\widehat{\mathbf{r}}}_2^c \right)} \nonumber 
\\ 
&& = U {\left[ {\left( 1 - D \right)} {\widehat{\mathbf{R}}}_1 {\widehat{\mathbf{A}}}_1 {\widehat{\mathbf{R}}}_1^c - D {\widehat{\mathbf{r}}}_2 {\widehat{\mathbf{A}}}_2 {\widehat{\mathbf{r}}}_2^c \right]} \nonumber 
\\ 
&& = U {\widehat{\mathbf{R}}}_1 {\widehat{\mathbf{A}}}_1 {\widehat{\mathbf{R}}}_1^c - U D {\widehat{\mathbf{G}}}_1. \label{BasTranMatA1}
\end{eqnarray}
In passing to the second row of the above multiple equation, we have used the first property in Eq. (\ref{SymplSProp1}) as well as Eq. (\ref{SymplRProp}). In a similar manner, we obtain 
\begin{eqnarray}
{\left( {\widehat{\mathbf{g}}}_1 + {\widehat{\mathbf{g}}}_2^c \right)} {\widehat{\mathbf{g}}}_2 && = U {\left( {\widehat{\mathbf{r}}}_1 {\widehat{\mathbf{R}}}_1^c {\widehat{\mathbf{R}}}_1 {\widehat{\mathbf{A}}}_1 {\widehat{\mathbf{r}}}_1^c - 
{\widehat{\mathbf{R}}}_2 {\widehat{\mathbf{r}}}_2^c {\widehat{\mathbf{r}}}_2 {\widehat{\mathbf{A}}}_2 {\widehat{\mathbf{R}}}_2^c \right)} \nonumber 
\\ 
&& = U {\left[ D {\widehat{\mathbf{r}}}_1 {\widehat{\mathbf{A}}}_1 {\widehat{\mathbf{r}}}_1^c - {\left( 1 - D \right)} {\widehat{\mathbf{R}}}_2 {\widehat{\mathbf{A}}}_2 {\widehat{\mathbf{R}}}_2^c \right]} \nonumber 
\\ 
&& = - U {\widehat{\mathbf{R}}}_2 {\widehat{\mathbf{A}}}_2 {\widehat{\mathbf{R}}}_2^c + U D {\widehat{\mathbf{G}}}_2. \label{BasTranMatA2}
\end{eqnarray}
This completes the derivation of Eqs. (\ref{MatrixA1}) and (\ref{MatrixA2}).



\bibliography{aipsamp}

\end{document}